\documentclass[mina4paper,11pt]{article}
\usepackage[width=15cm,height=23cm]{geometry} 
\usepackage[usenames]{color}
\usepackage{amsmath}
\usepackage{amsfonts,amssymb}
\usepackage{theorem}
\usepackage[dvips]{graphicx}

\newcommand{\dB}{\textbf{s}}
\newcommand{\Dintegral}[1]{\int \mathrm{d}^D{#1}}
\newcommand{\QA}{\Gamma}

\newcommand{\quoz}{\pi}
\newcommand{\subg}{\sigma}

\begin{document}

\vskip 3.0 truecm

\Large
\centerline{Nonlinear Realization of the $SU(5)$ Georgi-Glashow Model}
 
\normalsize
\rm 
\large
\vskip 1.1 truecm
\centerline{D.~Bettinelli\footnote{e-mail: {\tt daniele.bettinelli@mi.infn.it}},
R.~Ferrari\footnote{e-mail: {\tt ruggero.ferrari@mi.infn.it}},
A.~Sanzeni\footnote{e-mail: {\tt alessandro.sanzeni@gmail.com}}}
\normalsize
\medskip
\begin{center}
Dip. di Fisica, Universit\`a degli Studi di Milano\\
and INFN, Sez. di Milano\\
via Celoria 16, I-20133 Milano, Italy\\
(IFUM-1001-FT, October 2012)
\end{center}

\vskip 1.5 truecm

\centerline{\bf Abstract}

\rm
\begin{quotation}
The formulation of a grand unified field theory based on the nonlinearly realized $SU(5)$ gauge group is presented.
The tree-level action is constructed by requiring invariance under local left $SU(5)$ gauge transformations, 
the existence of a weak power-counting bound and by imposing also an additional invariance under local right 
$SU_C(3) \times U_Q(1)$ transformations. The local functional equation associated to this latter invariance allows one 
to prove the absence of physical scalar modes with the quantum numbers of massless gauge bosons from the perturbative spectrum.
Two independent mass invariants for the electroweak gauge bosons exist, although the gauge group of our model is simple.  
The flavour sector of the model is analyzed and compared with the one of the minimal Georgi-Glashow grand unified theory.  
\end{quotation}

\newpage

\section{Introduction}
\label{sect.intr}

Recently the formulation of gauge theories  with nonlinearly realized symmetry has been discussed in the framework of the 
electroweak model \cite{Bettinelli:2008ey,Bettinelli:2008qn,Bettinelli:2009wu}.
Even if  these kind of models are not power counting renormalizable, due to the
presence of non-polynomial interaction vertices in the Feynman rules,  
a consistent, i.e. symmetric, local and predictive theory  can be  defined in the loopwise expansion.
The  formulation of the theory relies on two pillars: the local functional equation (LFE) encoding the invariance 
of the path-integral measure under the non-linearly realized gauge symmetry 
and the weak power counting (WPC) criterion.
The LFE \cite{Ferrari:2005ii,Bettinelli:2007kc} uniquely fixes the dependence of the 1-PI amplitudes involving at least one Goldstone 
field (descendant amplitudes) in terms of 1-PI amplitudes with no external Goldstone legs (ancestor amplitudes), thus providing
a complete hierarchy among 1-PI Green's functions \cite{Ferrari:2005va}.
The WPC prescription \cite{Ferrari:2005va,Bettinelli:2007tq,Bettinelli:2008qn}, i.e. the request of having a finite number of divergent 
ancestor amplitudes order by order in perturbation theory, puts severe constraints on the allowed terms in the tree level action 
forbidding anomalous couplings.

A detailed analysis shows \cite{Bettinelli:2008ey,Bettinelli:2008qn} 
that besides the usual mass term for gauge bosons, another independent mass parameter can be added to the model. 
Thus, although the theory can account for the Weinberg relation, a fine-tuning is needed due to the extra mass term.
It has been conjectured that the appearance of the two independent mass invariants could be due to the abelian factor of the gauge group $SU(2)\times U(1)$.
Therefore, it can be interesting to study the case of a nonlinearly realized grand unified theory, in which the groups $SU(2)$ and $U(1)$ are both 
subgroups of a simple group. In the literature there are many models that can be used for this goal (see for instance Ref.~\cite{Langacker:1980}).
For the sake of simplicity we had chosen the smallest simple group, that is the one proposed by Georgi and Glashow \cite{Georgi:1974}, namely $SU(5)$.
In this context, we can also address another issue raised in the nonlinear electroweak model, i.e. the number of physical scalars present in the 
perturbative spectrum of the model. In particular,  the absence of physical scalar particles appears to be a consequence of simple counting of number of 
degrees of freedom (three massive gauge bosons and three scalar modes in the $SU(2)$-valued nonlinear sigma model field $\Omega$). 
This is not the case of the nonlinearly realized Georgi-Glashow model, in which the number of scalar modes present in the $SU(5)$-valued nonlinear sigma
model field is larger than the number of massive gauge bosons.

In this paper we derive the tree level action of a Georgi-Glashow model in which the gauge symmetry is nonlinearly realized in the scalar sector.
We show that the requirement of invariance under local left gauge transformations and the existence of a WPC bound 
do not fix completely the tree-level Feynman rules, since one is allowed to introduce many different mass terms.
An additional local right symmetry acting only on the scalar sector of the theory is the key ingredient in order to select a phenomenologically viable 
subset of mass terms. Indeed, we show that our model reduces in the low-energy limit to the nonlinearly realized $SU(2)\times U(1)$ electroweak theory.
Hence, though the gauge group we consider is simple,  we find two independent mass terms for the low mass gauge bosons $W$ and $Z$.
Furthermore, the right symmetry of the grand unified scalar sector is hidden. By this we mean that at low energies all the mass terms can be expressed 
in terms of composite fields which are invariant under local right transformations.

As a result of the local right invariance of the tree level action, one can remove the scalars associated to massless gauge bosons.
This symmetry allows us to derive a further LFE (right) satisfied  by the tree level action. 
By imposing that this equation holds also for the full quantum action, 
we prove that all physical S-matrix elements involving at least  one scalar mode associated to a massless gauge boson are identically zero. 
This entails the absence of physical scalars from the perturbative spectrum of the model.

As a phenomenological application of our formalism we have performed a comparison between the flavor sector of our model and the one of Georgi-Glashow.
It turns out that the nonlinear formulation poses less stringent constraints in the relations between fermion masses at the grand unified scale.
Moreover, the presence of more  independent mixing matrices than in the minimal Georgi-Glashow model allows one to  increase the proton lifetime, 
by means of a fine-tuning of the free parameters.

The paper is organised as follows.
In Sec.~\ref{sect.lagr} we introduce the symmetries of the model and derive the tree level action.
Sec.~\ref{sect.flav} deals with the flavor sector of our model.
In Sec.~\ref{sect.phys} we show how to remove the physical scalars from the tree level Feynman rules.
The 1-PI generating functional is derived in Sec.~\ref{sec.PI} and used to obtain LFEs for the left and 
the right symmetries in Sec.~\ref{sect.LFE}. We also show in simple examples how to use the right equation to prove the decoupling of 
physical scalars from the perturbative spectrum.
The WPC bound is formulated in Sec.~\ref{sect.wpc}, while in Sec.~\ref{sect.con} we give our conclusions and outlook.
The appendices contain our notation, a proof of the nilpotency of the BRST transformations on the scalar fields and a 
derivation of the low-energy limit of the model.

\section{Lagrangian of the model}
\label{sect.lagr}

In this section we derive the Lagrangian of the nonlinearly realized $SU(5)$ model.
The kinetic and the interaction terms of fermions and gauge bosons coincide with 
those of the Georgi-Glashow Lagrangian \cite{Georgi:1974}.
Therefore here we discuss only  mass terms for matter and gauge fields.
As  anticipated in the introduction, we use terms \textit{\`a la} St\"uckelberg instead of the Higgs 
mechanism in order to give mass to the fields.
A local right symmetry turns out to be the key ingredient to select the Lagrangian of the theory.
Indeed, we show (see Appendix~\ref{app.3}) that our model reduces in the low-energy limit to the nonlinearly realized $SU(2)\times U(1)$ electroweak theory.
Hence, two independent mass terms for the electroweak gauge bosons $W$ and $Z$ exist, although the gauge group of the original model is simple.

\subsection{Gauge mass terms}
\label{sect.gaugem}

We define the bleached gauge fields \cite{Bettinelli:2007tq} and use them in order to construct mass terms 
for the gauge bosons. By requiring only local invariance under $SU(5)$ left transformations one can show that 
there are $24$ independent mass terms. A phenomenologically viable model is obtained by imposing 
also a suitable local right symmetry.

We consider the gauge connection $A_{\mu}=A_{\mu}^a L^a$, where $L^a$ are the generators of 
the $SU(5)$ group in the regular representation. The nonlinear sigma model field $\Omega$ is an element of the $SU(5)$ group.
An explicit parametrization in terms of scalar fields is not needed in this section.
The $SU(5)$ flat connection $F_{\mu}$ is defined in terms of $\Omega$ according to 
\begin{equation}
F_{\mu}(x) =F_{\mu}^a(x) L^a = i \Omega(x) \partial_\mu \Omega^{\dagger}(x) \, .
\label{eq.1}
\end{equation}
Under local $SU(5)$ left transformations, the fields transform according to
\begin{eqnarray}
&&
A_{\mu}  \rightarrow U A_{\mu} U^{\dagger}+\frac{i}{g_5} U \partial_{\mu}U^{\dagger}\,,~~~~~
\Omega   \rightarrow   U\Omega\,,
\nonumber\\ 
&& F_{\mu}  \rightarrow U F_{\mu} U^{\dagger}+i U \partial_{\mu}U^{\dagger}\,,
\label{eq.2}
\end{eqnarray}
where $g_5$ is the gauge coupling constant and $U$ is a $SU(5)$ matrix.
One can construct out of $A_\mu -F_\mu$ and $\Omega$ a bleached gauge field
\begin{equation}
a_{\mu} = a_{\mu}^b L^b := {\Omega}^{\dagger} \left( g_5 A^{\mu} - F^{\mu}  \right) {\Omega}
=g_5 {\Omega}^{\dagger} A^{\mu}  {\Omega} + i \Omega^{\dagger} \partial_\mu \Omega\,.
\label{eq.3}
\end{equation}
We point out that every component $a^i_j := {\rm Tr}\big(a_\mu t^i_j\big)$, where the matrices $t^i_j$ are defined in Appendix~\ref{app.1}, 
of $a_{\mu}$ is separately invariant under local gauge transformations (\ref{eq.2}).
Thus, we find the following quadratic invariants for any $M^{jk}_{il}$
\begin{equation}
M^{jk}_{il} a_j^i\, a_k^l \, .
\label{eq.4}
\end{equation}
It is possible to define also a set of local right transformations 
\begin{eqnarray}
&&
A_{\mu}  \rightarrow  A_{\mu}\,, ~~~~~~~~~~~~~~~~~~~~~~~~~ 
\Omega  \rightarrow  \Omega\, V^{\dagger} \,,\nonumber\\
&& 
F_{\mu} \rightarrow   F_{\mu} +i \Omega V^{\dagger} (\partial_{\mu} V ) \Omega^{\dagger}\,,~~~~
a_{\mu} \rightarrow  V a_{\mu} V^{\dagger}+i V \partial_{\mu}V^{\dagger} \,,
\label{eq.5}
\end{eqnarray}
with $V$ belonging to a subgroup of $SU(5)$.
The invariance of the action under the transformations in eq.(\ref{eq.5}) will turn 
out to be crucial in order to discriminate among different $SU(5)$ grand unified models.  
Both the local left and right symmetries are  spontaneously broken by the vacuum expectation 
value (VEV) of  $\Omega$. 
One can always assume without loss of generality that the VEV is given by
\begin{equation}
\langle \Omega \rangle = \Omega_0 = \mathbb{I}\, .
\label{eq.6}
\end{equation}
This leaves unbroken the generators $Q$ of a vectorial transformation, $\Omega \to V \Omega V^\dagger$, which act 
on $\Omega$ as 
\begin{equation}
\delta\Omega =i [Q,\Omega]   \, .
\label{eq.7}
\end{equation}
From the above equation we see that the charge operators, that can be used to label 
the states of the system, are in a one-to-one correspondence with the generators of the right symmetry.
Phenomenology of particle physics requires an exact $SU_C(3)\times U_Q(1)$ gauge symmetry, 
i.e. color and electric charge symmetry.
This implies that $V$ belongs to the $SU_C(3)\times U_Q(1)$ subgroup of $SU(5)$ 
(see appendix~\ref{app.1} for the notations).
Since $a_{\mu}$ behaves as a gauge connection under the right transformations (\ref{eq.5}),  among the quadratic terms in 
equation (\ref{eq.4}) those which are also right invariant are
\begin{equation}
\label{eq.8}
M_1^2\,    a_{t}^{\alpha}  a^{t}_{\alpha}
+M_2^2\,   a^{5}_{\alpha}  a_{5}^{\alpha} 
+M_3^2\,   a^{t}_{5}  a^{5}_{t}
+M_4^2\,  a^5_5  a^5_5  
\end{equation}
with $\alpha=1,2,3$ and $t=4,5$.\\

\subsection{Fermion mass terms}
\label{sect.ferm}

We assign fermions to the $SU(5)$ representations used by Georgi and Glashow \cite{Georgi:1974}, i.e. 
the $\bar{5}$ $\psi^i$ and the $10$ $\chi_{ij}$ (see appendix~\ref{app.1} for the explicit expressions).
They transform as 
\begin{equation}\begin{split} 
&\psi^i \rightarrow	  	\psi^j\, (U^\dagger)^i_j \,, 				\\
&\chi_{ij} \rightarrow		U^k_i\, U^l_j\, \chi_{kl} \, .
\label{eq.9}
\end{split}\end{equation}
under  local $SU(5)$ left transformations.
Both $\psi$ and $\chi$ are right invariant.
The $SU(5)$-valued field $\Omega$ is used to construct also fermion bleached fields, 
\begin{equation}
\begin{split} 
&\widetilde{\psi}^i :=  	\psi^j\, \Omega_j^i \,,				\\
&\widetilde{\chi}_{ij}:= (\Omega^{\dagger})_i^k\, (\Omega^{\dagger})_j^l \, \chi_{kl}\,, 
\end{split}
\label{eq.10}
\end{equation}
which are gauge invariant and transform as 
\begin{equation}
\begin{split} 
&\widetilde{\psi}^i \rightarrow \widetilde{\psi}^j\, (V^{\dagger})_j^i \,,			\\
&\widetilde{\chi}_{ij} \rightarrow V_i^k\, V_j^l\, \chi_{kl} 
\end{split}
\label{eq.11}
\end{equation}
under right transformations.

Since we are considering a theory that has $SU_C(3)\times U_Q(1)$  right symmetry  the invariant 
quadratic terms are
\footnote{There is the possibility to add also another term $(\widetilde{\psi}^T)^5\, \mathcal{C}\,
\widetilde{\psi}^5$ which gives a Majorana mass to the neutrinos. However, we do not include this 
term in our Lagrangian.}
\begin{eqnarray}
&&
(\widetilde{\psi}^{T})^\alpha \,\mathcal{C}\,  \widetilde{\chi}_{\alpha 5} + \textit{h.c.}\,,~~~~~~~~~~
(\widetilde{\psi}^T)^{4} \,\mathcal{C}\,  \widetilde{\chi}_{4 5} + \textit{h.c.}\,,	\nonumber\\
&&
\epsilon^{5ijkl}(\widetilde{\chi}^T)_{ij}\, \mathcal{C}\, \widetilde{\chi}_{kl} + \textit{h.c.}\,,
\label{eq.12}
\end{eqnarray}
where $\mathcal{C}$ is the usual charge conjugation matrix. 

We point out that the fifth index of every multiplet is 
not contracted since it is left 
unchanged  by  the right transformations. 
Indeed, by studying the generators of the right matrix $V$ given in appendix~\ref{app.1},  one can see that $V_5^5=1$ and $V_4^5=V_5^4=0$.

It is rather interesting to notice that 
if one requires invariance under $SU(5)$ right symmetry, instead of $SU(3)_C \times U(1)_Q$,  
the model under consideration does not admit fermion mass terms.

If we consider more families, using  equation (\ref{eq.12}) we have that  the mass terms 
of fermion fields in the 
Lagrangian are:
\begin{eqnarray}
\label{eq.13}
-M^d_{kn} \, (\widetilde{\psi}_k^T)^{\alpha}\, \mathcal{C}\,  (\widetilde{\chi}_{n})_{\alpha 5} 
-M^e_{kn} (\widetilde{\psi}_k^T)^{4}\, \mathcal{C}\,  (\widetilde{\chi}_{n})_{4 5}
-M^u_{kn} \epsilon^{5ijhl} (\widetilde{\chi}_k^T)_{ij} \,\mathcal{C}\, (\widetilde{\chi}_{n})_{hl}
+\textit{h.c.} 
\end{eqnarray}
where $k,n$ are family indexes. 
Since $(\widetilde{\chi}_k^T)_{ij}\, \mathcal{C}\, (\widetilde{\chi}_{n})_{hl}
 = (\widetilde{\chi}_n^T)_{hl}\, \mathcal{C} \,(\widetilde{\chi}_{k})_{ij}$, 
we can take the matrix $M^u$ symmetric.

\subsection{Lagrangian in the gauge eigenstate basis}
\label{sect.2.3}

The above discussion uniquely fixes 
the gauge-invariant part of the classical action of our model
\begin{eqnarray}
\!\!\!\!\!\!\!\!\!\!\!\!\!\!
S_0 &\!\!\! = &\!\!\! \frac{\Lambda^{(D-4)}}{g_5^2} \int \! d^Dx\,\Big[
i\bar{\psi} \, \!\not\!\!D \psi + i\bar{\chi} \, \!\not\!\!D \chi -\frac{1}{2} 
{\rm Tr}\big(G_{\mu \nu}[A] G^{\mu \nu}[A]\big) 
\nonumber\\
&&~~~~~~~~~~~~~~~~
+M_1^2\,    a_{t}^{\alpha}  a^{t}_{\alpha}
+M_2^2\,   a^{5}_{\alpha}  a_{5}^{\alpha} 
+M_3^2\,   a^{t}_{5}  a^{5}_{t}
+M_4^2\,  a^5_5  a^5_5 \nonumber\\
&&\!\!\!\!\!
-M^d_{kn} (\widetilde{\psi}_k^{T})^{\alpha}\, \mathcal{C}\,  (\widetilde{\chi}_n)_{\alpha 5} 
-M^e_{kn} (\widetilde{\psi}_k^{T})^{4}\, \mathcal{C} \, (\widetilde{\chi}_n)_{4 5}
-M^u_{kn} \epsilon^{5ijhl} (\widetilde{\chi}_k^T)_{ij}\, \mathcal{C} \,(\widetilde{\chi}_n)_{hl}
+\textit{h.c.}\Big]\,,
\label{eq.2.14}
\end{eqnarray}
where $\Lambda$ is a mass scale for continuation 
in the dimensions, while the gauge covariant derivatives $D_\mu$ are defined by
\begin{eqnarray}
\label{eq:kin}
&&
\left( D_\mu \psi \right)^a= \partial_\mu \psi^a-i  \frac{g_5}{\sqrt{2}} (A_\mu )_b^a  \psi^b \,, \nonumber\\
&&
\left( D_\mu \chi \right)_{ab} = \partial_\mu \chi_{ab}+ i \frac{g_5}{\sqrt{2}} (A_\mu )_a^c \chi_{cb}
+ i \frac{g_5}{\sqrt{2}} (A_\mu )_b^d \chi_{ad} 
\end{eqnarray}
and the field strength $G_{\mu \nu}[A] $ is given by 
\begin{equation}
\left(G_{\mu \nu}[A]\right)_a^b=\partial_{\mu} \left( {A}_{\nu}   \right)_a^b 
-\frac {ig_{5}}{\sqrt 2} \left( {A}_{\mu}\right)_a^c  \left({A}_{\nu} \right)_c^b
-\left( \mu \leftrightarrow \nu\right) \, .
\end{equation}
For phenomenological reasons, one assumes that the $X$ and $Y$ gauge bosons are extremely heavy compared to the other particles. 
This can be achieved by imposing a hierarchy among mass parameters, i.e. $M_1 \gg M_2,M_3,M_4,M^d_{kn},M^u_{kn},M^e_{kn}$. One can prove 
(see Appendix~\ref{app.3}) that in the low-energy limit the mass terms in the second and third line of the action in eq.(\ref{eq.2.14}) 
reduce to the ones of the nonlinearly realized $SU(2)\times U(1)$ electroweak model \cite{Bettinelli:2008ey, Bettinelli:2008qn}.

\section{Lagrangian in the mass eigenstate basis and CKM matrix}
\label{sect.flav}

In this section we introduce the mass eigenstate basis for the matter and gauge bosons fields. 
In this way we identify the photon and the $Z$ boson in the tree-level action.
We will show that there are two independent mass terms for the $W$ and $Z$ gauge bosons 
even though the gauge group is simple.
Furthermore, we find that the flavor sector of our model is less constrained than the one of the Georgi-Glashow model. 
This might have an impact on the  proton lifetime in the nonlinear model.

The mass term of $A_{\mu}$ can be derived from the one of $a_{\mu}$ by substituting $\Omega$ with its VEV.
Since $\Omega_0= \mathbb{I}$, one sees that  $(A_{\mu})_n^m$ and $(a_{\mu})_n^m$ have the same mass parameter. 
Starting from eq.(\ref{eq.2.14}) we notice that there are some combinations of the gauge fields $A^a_\mu$ which are massless. 
In particular, in the neutral electroweak sector,  the combination given by
\begin{equation}
{\rm Tr}\left[ A_{\mu} L^{\gamma} \right] \,,~~ {\rm where}~~ L^{\gamma} = \sqrt{\frac{3}{8}}\, L_{11} +\sqrt{\frac{5}{8}}\,
L_{12}   
\label{eq.3.3}
\end{equation}
has the quantum numbers of the photon.
We define the matrix $L^{Z}$ as the linear combination of $L_{11}$ and $L_{12}$ orthogonal to $L^{\gamma}$,
\begin{equation}
L^{Z}:=\sqrt{\frac 58} L^{11} -\sqrt{\frac 38} L^{12} \,.
\end{equation}
With these definitions the new generators are orthogonal and correctly normalized, indeed
\begin{equation}
{\rm Tr}\left(L^{\gamma} L^{Z}\right) = 0\,,~~~ 
{\rm Tr}\left(L^{\gamma} L^{\gamma}\right) = {\rm Tr}\left(L^{Z} L^{Z}\right) = \frac 12 \, .
\label{eq.3.5}
\end{equation}
The above specified  change of basis can be expressed with a rotation
\begin{equation}\label{eq.wwwww}
\left(
\begin{array}{cc}
L^{Z} \\
L^{\gamma} 
\end{array}
\right)=
\left(
\begin{array}{cc}
\cos(\theta_w) & -\sin(\theta_w) \\
\sin(\theta_w) & \cos(\theta_w) 
\end{array}
\right) \
\left(
\begin{array}{cc}
L^{11} \\
L^{12} 
\end{array}
\right)
\end{equation}
where $\theta_w$ is the Weinberg angle and $\sin^2(\theta_w)=\frac 38$. 
This tree-level value, valid only at the unification scale, coincides with the one obtained in the linear Georgi-Glashow model \cite{Georgi:1974}.
In order to compute the Weinberg angle at the electroweak scale, one has to consider the low-energy limit of the nonlinear $SU(5)$ model 
derived in Appendix~\ref{app.3}. Since the electroweak part of the low-energy Lagrangian coincides with the one considered in Refs.~\cite{Bettinelli:2008qn}, 
one can use the all-orders definition of the photon and $Z$ boson fields given in that paper. This gives an explicit prescription in order to define the 
rotation matrix and the Weinberg angle order by order in perturbation theory.
According to our conventions, in the Landau gauge the all-order Weinberg rotation is given by
\begin{equation}\label{loopWeinb}
\left(
\begin{array}{cc}
L^{Z} \\
L^{\gamma} 
\end{array}
\right)=
\frac1{\sqrt{\left( \Gamma^{BB}_L\right)^2+\left( \Gamma^{BA}_L\right)^2}}
\left(
\begin{array}{cc}
 \Gamma^{AB}_L\ &  \Gamma^{BB}_L\ \\
 -\Gamma^{BB}_L\ &  \Gamma^{AB}_L\ 
\end{array}
\right) \
\left(
\begin{array}{cc}
L^{11} \\
L^{12} 
\end{array}
\right)
\end{equation}
where the subscript $L$ stands for the longitudinal part of the corresponding 1-PI Green's function.
We expect to find sizable corrections to $\sin^2(\theta_w)$ because of the presence of large logarithms of the mass ratio $M_X/M_W$.   
 
Moreover, all of the gauge bosons belonging to the subgroup $SU(3)$ of $SU(5)$ are massless and 
correspond the gluons. 

In the new basis we see that the two mass terms with parameters $M_3$ and $M_4$ in eq.(\ref{eq.2.14}) 
give independent mass to the $Z$ and $W$ bosons.
A straightforward analysis of the gauge bosons mass Lagrangian gives 
\begin{equation}
M_X^2=  M_1^2\,,~~~
M_Y^2= M_1^2  +  M_2^2\,,~~~  
M_W^2=M_3^2 \,,~~~	
M_Z^2=\frac {8}{5} \left( M_3^2 +M_4^2 \right)  \, .
\label{eq.3.6}
\end{equation}

In order to obtain the mass eigenstate basis for the matter fields we have to 
diagonalize the matrices $M^d$, $M^u$ and $M^e$ of equation (\ref{eq.2.14}).
As usual, this can be done by means of a biunitary transformation, 
i.e. there exist unitary matrices $A_{L,R}^{u,d,e}$ such that
\begin{equation}
{A_{L}^{d}}^{\dagger} \ M^d \ {A_{R}^{d}} = M^d_D \ ,
\label{eq.3.7}
\end{equation}
where $M^d_D$ is diagonal and similarly for $M^e$ and $M^u$.
These matrices define the following change of basis
\begin{equation}
\left(d'_{L}\right)_n = \left(A_{L}^{d}\right)_{nk} \left(d_{L}\right)_k \,,~~~~
\left(d'_{R}\right)_n = \left(A_{R}^{d}\right)_{nk} \left(d_{R}\right)_k\,,
\label{eq.3.8}
\end{equation}
between the gauge eigenstate $d$ and the mass eigenstate $d'$.
In the same way one can define also the change of basis between the gauge eigenstates 
$u$, $e$ and the corresponding mass eigenstates $u'$, $e'$.
 
In the new basis the fermion  mass part of the Lagrangian becomes
\begin{equation}
- \bar u'_L M^u_D u'_R - \bar d'_L M^d_D d'_R -\bar e'_L M^e_D e'_R+\textit{h.c.}
\label{eq.3.9}
\end{equation}
Furthermore, the interaction Lagrangian between fermions and the $X$ and $Y$ gauge bosons, i.e.  
\begin{eqnarray}
&&
\frac{ g_5}{\sqrt 2}  \left[ \bar{e} _{R}^{+}  \not \!{X}^\alpha (d_{R})_{\alpha} 
+\bar{e} _{L}^{+}  \not \!{X}^\alpha (d_{L})_{\alpha} 
- \bar{\nu}^c_R \not \!{Y}^\alpha (d_{R})_{\alpha} 
- \bar{e} _{L}^{+}  \not \!{Y}^\alpha (u_{L})_{\alpha} \right] \nonumber\\
&&\!\!\!
-  \frac{ g_5}{\sqrt 2}  \left[ \epsilon ^{\alpha \beta \gamma}(\bar{u}^c_{L})_{ \gamma }  
\not \! \bar{X}_\alpha (u_{L})_{\beta}
+  \epsilon^{\alpha \beta \gamma}(\bar{u}^c_{L})_{ \gamma}  \not \! \bar{Y}_\alpha 
(d_{L})_{\beta}\right] + \text{h.c.}	\, ,
\label{eq.3.10}
\end{eqnarray}
can be rewritten in the mass eigenstate basis
\begin{eqnarray}
&&\!\!\!\!\!\!\!
\frac{g_5}{\sqrt 2}  \Big[
\left( (A_{R}^{e})^{\dagger} A_{R}^{d} \right)_{mn} \bar{e'}_{mR}^{+}  \not \!{X}^\alpha (d'_{nR})_{\alpha} 
+\left( (A_{L}^{e})^{\dagger} A_{L}^{d} \right)_{mn} \bar{e'}_{Lm}^{+}  \not \!{X}^\alpha (d'_{nL})_{\alpha}
\nonumber\\
&& ~~
- \left( (A_{R}^{e})^{\dagger} A_{R}^{d} \right)_{mn} \bar{\nu'}^c_{mR }\not \!{Y}^\alpha (d'_{nR})_{\alpha}
 - \left( (A_{L}^{e})^{\dagger} A_{L}^{u} \right)_{mn} \bar{e'}_{Lm}^{+}  \not \!{Y}^\alpha (u'_{Ln})_{\alpha} 
\Big] \nonumber \\
&&\!\!\!\!\!\!\!\!\!\! 
-  \frac{g_5}{\sqrt 2}  \left[   
 K_{mn}  \epsilon ^{\alpha \beta \gamma}(\bar{u'}^c_{Lm})_{ \gamma }  \not \! \bar{X}_\alpha (u'_{Ln})_{\beta}
+ \left(K {A_C}^{\dagger} \right)_{mn} \epsilon^{\alpha \beta \gamma}(\bar{u'}^c_{L m})_{\gamma}  
\not \! \bar{Y}_\alpha (d'_{Ln})_{\beta}\right] 
+ \text{h.c.}	\,.
\label{eq.3.11}
\end{eqnarray}
Here $A_C= (A_L^d)^\dagger A_L^u$ is the CKM matrix, while $K$ is a diagonal matrix of phases such that 
 $\big(A_{R}^{u}\big)^*=A_{L}^{u} K^*$. It is uniquely determined by the request that $M_D^u$ be real and positive. 
We remark that in the change of basis also the electroweak interactions of quarks get modified.
However, the latter has not been reported here because the expressions in both basis coincide 
with the same quantities in the Standard Model.

Some comments are in order. 
i) In the minimal linear Georgi-Glashow model  one finds\cite{Langacker:1980} $M^d=M^e$ which implies the following relation
\begin{equation}
\frac{m_d}{m_s} = \frac{m_{e}}{m_{\mu}} \ ,
\label{eq.3.12}
\end{equation}
valid also at low-energy scales because of its renormalization group invariance.
This is in strong disagreement with observations.
At variance with this, in our theory we find that $M^d$ and $M^e$ are completely independent, 
hence we do not have this problem.
ii) We find that the di-quark  and lepto-quark interaction vertices in our model are different  
from those of the minimal Georgi-Glashow model \cite{Langacker:1980}, as one can see by comparing equation (\ref{eq.3.11}) 
with the corresponding term of the linear  model, i.e. 
\begin{eqnarray}
&&
\frac{ g_5}{\sqrt 2}  \big[
 \bar{e'} _{R}^{+}  \not \!{X}^\alpha (d'_{nR})_{\alpha} 
+\bar{e'} _{Ln}^{+}  \not \!{X}^\alpha (d'_{nL})_{\alpha} 
-\bar{\nu'}^c_{nR }\not \!{Y}^\alpha (d'_{nR})_{\alpha}
- \left( {A_C} \right)_{mn}\bar{e'} _{Lm}^{+}  \not \!{Y}^\alpha (u'_{Ln})_{\alpha} 
\big]\nonumber \\
&&\!\!\!
-  \frac{ g_5}{\sqrt 2}  \big[   
 K_{mn}  \epsilon ^{\alpha \beta \gamma}(\bar{u'}^c_{Lm})_{ \gamma }  \not \! \bar{X}_\alpha (u'_{Ln})_{\beta}
+ \left(K {A_C}^{\dagger} \right)_{mn} \epsilon^{\alpha \beta \gamma}(\bar{u'}^c_{L m})_{\gamma}  
\not \! \bar{Y}_\alpha (d'_{Ln})_{\beta}\big] 
+ \text{h.c.}	\, .
\label{eq.3.13}
\end{eqnarray}
iii) As shown in Ref.~\cite{Nandi:1982ew} in our model the proton decay cannot be rotated away with a particular choice of the parameters of the mixing matrices.
However, as pointed out in Ref.~\cite{Mohapatra:1979yj}, the proton life-time can be rendered as long as one wishes by introducing a fine-tuning 
of parameters of the mixing matrices.

\section{Absence of physical scalar bosons in the perturbative spectrum}
\label{sect.phys}

In the nonlinearly realized  $SU(5)$ model there are $24$ independent scalar fields associated to $\Omega$, each of which 
is in one-to-one correspondence with a gauge field having the same quantum numbers.
In this section we shall prove that those scalars associated to the gluons and the photon do not appear in 
the  tree-level Feynman rules.
This implies that our model can be  consistently formulated on the quotient space $SU(5)/ \left(SU_C(3) 
\times U_Q(1) \right)$.

First of all we introduce a parametrization of the $SU(N)$-valued field, which was devised 
by Coleman et al.~\cite{Coleman}.\\
Let $L_a$ be a complete set of generators of $SU(5)$, $a=1, \dots 24 $  (see appendix~\ref{app.1} for the explicit expressions).
We call $S_{\dot{a}}$ ($\dot{a}=1, \dots 9 $) the generators of the  $SU_C(3) \times U_{Q}(1)$ sub-group and $P_{\ddot{a}}$ ($\ddot{a} = 10, \dots 24 $) 
the remaining elements of the algebra $su(5)$.
With this definition, each element $U$ of $SU(5)$ admits the unique decomposition:
\begin{equation}\label{ColemanPar}
U = \exp\big(\alpha_{\ddot{a}} P_{\ddot{a}}\big)\, \exp\big(\beta_{\dot{a}} S_{\dot{a}}\big) \, .
\end{equation}
For every element $w \in SU(5)$, one has: 
\begin{equation}
w \, \exp\big(\alpha_{\ddot{a}} P_{\ddot{a}}\big)\, \exp\big(\beta_{\dot{a}} S_{\dot{a}}\big) =   
\exp\big(\alpha'_{\ddot{a}} P_{\ddot{a}}\big)\, \exp\big(\beta'_{\dot{a}} S_{\dot{a}}\big)
\end{equation}
with
\begin{equation}
 \alpha' = \alpha'\left( \alpha,\beta,w \right)\,,~~~
 \beta' = \beta'\left( \alpha,\beta,w \right) \, .
\end{equation}
Furthermore, for $h \in SU_C(3)\times U_Q(1)$, one gets:
\begin{equation}
\exp\big(\alpha_{\ddot{a}} P_{\ddot{a}}\big)\, \exp\big(\beta_{\dot{a}} S_{\dot{a}}\big)\, h =   
\exp\big(\alpha_{\ddot{a}} P_{\ddot{a}}\big)\, \exp\big(\beta'_{\dot{a}} S_{\dot{a}}\big) 
\end{equation}
with
\begin{equation}
 \beta' = \beta'\left(\beta,h \right) \, .
\end{equation}

From now on, we will use the Coleman parametrization to express the $SU(5)$-valued field $\Omega$
\begin{equation}\label{OmegaColeman}
\Omega = \exp\big(\pi_{\ddot{a}} P_{\ddot{a}}\big)\, \exp\big(\sigma_{\dot{a}} S_{\dot{a}}\big) = \Pi\, \Sigma \,,
\end{equation}
where $\Sigma\in SU_C(3) \times U_Q(1)$ and $\Pi$ is an element of $SU(5)$ generated by those $L_a$ that do not belong 
to the Lie algebra of $SU_C(3) \times U_Q(1)$.
In this way $\Omega$ has been expressed in terms of $24$ scalars, namely $\subg_{\dot{a}}$ with ${\dot{a}} \in \{ 1, \dots 9\}$ 
and $\quoz_{\ddot{a}}$ with $\ddot{a} \in \{ 10, \dots 24 \}$.
It then follows that
\begin{equation}
a_{\mu} = \Sigma^{\dagger} b_{\mu} \Sigma + i  \Sigma^{\dagger} \partial_{\mu} \Sigma \, ,
\end{equation}
with
\begin{equation}
b_{\mu} = \Pi^{\dagger} \left(A_{\mu} -i \Pi \partial_{\mu } \Pi^{\dagger} \right) \Pi \, .  
\end{equation}
Hence, $a_{\mu}$ can be seen as the result of a local right $SU_C(3) \times U_Q(1)$ transformation acting on $b_\mu$, which, by construction, does 
not depend on the fields $\sigma_{\dot{a}}$.
The same procedure can be applied to the fermion fields.
Since the action (\ref{eq.2.14}) is invariant under $SU_C(3) \times U_Q(1)$ right transformations,  we can conclude that 
$$S_0(a_{\mu}) = S_0(b_{\mu})\,.$$
The above results show that  one can always eliminate 
the scalar fields associated to massless gauge bosons from the tree-level Feynman rules.
In section~\ref{sect.LFE} it will be shown that these fields can be consistently removed order by order in the loop expansion.

\section{Path integral formulation}
\label{sec.PI}

In this section we gauge-fix the classical action (\ref{eq.2.14}) and introduce the generating functional of Green's functions 
which will be used in the quantization of our model. The gauge-fixing is performed by BRST techniques in the Landau gauge for the sake of simplicity. 
The BRST differential $\dB$  is obtained in the usual way by promoting the gauge parameters $\omega^L_a$ of the local left transformations 
to the ghost fields $c_a$ and by introducing the antighosts $\bar{c}_a$ coupled in a BRST doublet to the Nakanishi-Lautrup 
fields, $b_a$.

The infinitesimal variations of the gauge and matter fields under local left transformations can be easily 
 obtained from eqs.(\ref{eq.2}), (\ref{eq.9}). 

Here we limit ourselves to the derivation of the transformation properties of the scalar fields $\quoz_{\ddot{a}}$ and $\subg_{\dot{a}}$, 
introduced in eq.(\ref{OmegaColeman}).
For local left transformations we know that $\Omega$ behaves as $ \Omega \rightarrow U \Omega $, 
where $U = \exp\big(\alpha^L_{\ddot{a}} P_{\ddot{a}}\big)\, \exp\big(\beta^L_{\dot{a}} S_{\dot{a}}\big)$ is a $SU(5)$ matrix.
Using the Baker-Campbell-Hausdorff formula and keeping only terms of the first order in $\alpha^L$ and $\beta^L$, 
we find that the infinitesimal left transformation properties of the scalar fields can be written as
\begin{eqnarray}
\label{variationscal}
&&
\delta^L\subg_{\dot{a}} =\left( \Theta^{(1,0)} \left( \quoz , \subg \right) \right)_{{\dot{a}}b} \omega^L_b \,,\nonumber\\
&&
\delta^L\quoz_{\ddot{a}} =\left( \Theta^{(1,0)} \left( \quoz , \subg \right) \right)_{\ddot{a}b} \omega^L_b \,,
\end{eqnarray}
where $\omega^L$ is a vector of components $(\beta^L, \alpha^L)$, while $ \Theta^{(1,0)}$ is a complicated function of the fields, 
whose explicit expression is not needed in what follows.
We remark that after a local left transformation the field $\Omega$ reads 
\begin{equation}
\exp\Bigg(\Big(\pi_{\ddot{a}}+ \left( \Theta^{(1,0)} \left( \quoz , \subg \right) \right)_{\ddot{a}b} \omega^L_b \Big) 
P_{\ddot{a}}\Bigg)\, \exp\Bigg(\Big(\sigma_{\dot{a}}+ \left( \Theta^{(1,0)} 
\left( \quoz , \subg \right) \right)_{{\dot{a}}b} \omega^L_b \Big)    S_{\dot{a}}\Bigg)
\end{equation}
so the $\sigma$ fields cannot be removed from the action as we have done in section \ref{sect.phys}.
However we can express the action in terms of new scalar fields 
\begin{equation}\begin{split}
\tilde{\pi}_{\ddot{a}}=\pi_{\ddot{a}}+ \left( \Theta^{(1,0)} \left( \quoz , \subg \right) \right)_{\ddot{a}b} \omega^L_b  \\
\tilde{\sigma}_{\dot{a}} =\sigma_{\dot{a}}+ \left( \Theta^{(1,0)} \left( \quoz , \subg \right) \right)_{{\dot{a}}b} \omega^L_b,
\end{split}\end{equation}
in this way  $\Omega$ can be cast in the standard form (\ref{OmegaColeman}) and the new field $\tilde{\subg}$ can be eliminated.

Under local right transformations the $SU(5)$-valued field transforms as $ \Omega \rightarrow  \Omega V^{\dagger}$, 
with $V = \exp\big(i \beta^R_{\dot{a}} S_{\dot{a}}\big)$. 
Thus, one finds
\begin{eqnarray}
\label{eq:lfer3}
&&
\delta^R \subg_{\dot{a}} =\left( \Theta^{(0,1)} \left(  \subg \right) \right)_{{\dot{a}}{\dot{b}}} \beta^R_{\dot{b}} \,,\nonumber\\
&&
\delta^R \quoz_{\ddot{a}} = 0\,.
\end{eqnarray}
Furthermore, thanks to the associativity of $SU(5)$, one can prove that
\begin{eqnarray}
\label{eq:lfer4}
\left[ \delta^L,\delta^R \right] = 0 \, .
\end{eqnarray}

The BRST differential of all the fields that appear in the gauge invariant action $S_0$ is given by
\begin{eqnarray}
\label{eq.63}
&& \dB \psi_{i} = i c_a \big(L^a\big)_i^j \psi_{j} \,,~~~~~~~\!  
   \dB \chi_{ij} = i c_a \big[\big(L^a\big)_i^k \chi_{kj} +\big(L^a\big)_j^k \chi_{ik}\big]\,, \nonumber\\
&& \dB A_{a\mu} = \left( D_{\mu} \left[ A\right] c \right)_a \, , ~~~~
   \dB \bar{c}_a  = b_a \, ,  \nonumber\\
&& \dB \subg_{\dot{a}} = \left( \Theta^{(1,0)}\right)_{{\dot{a}}b}\, c_b \, , ~~~~
   \dB \quoz_{\ddot{a}} = \left( \Theta^{(1,0)}\right)_{\ddot{a}b}\, c_b \,,  \nonumber\\
&& \dB b_a  = 0 \, .
\label{eq.brst}
\end{eqnarray}
The BRST transformation of $c_a$ follows by nilpotency
\begin{equation}\label{eq.64}
\dB c_a =-\frac 12 f_{abc} c_b c_c \, .
\end{equation}
In appendix~\ref{app.2} we prove that the action of $\dB$ on the scalar fields is nilpotent.

The gauge fixing part of the action is 
\begin{eqnarray}
\label{Eq:gf}
S_{gf} = \frac{\Lambda^{(D-4)}}{g_5^2}\,\, \dB \Dintegral{x}\, \Big(\bar{c}_a \partial_{\mu} A^{\mu}_a \Big) = 
\frac{\Lambda^{(D-4)}}{g_5^2} \Dintegral{x}\, \Big[b_a \partial_{\mu} A^{\mu}_a - \bar{c}_a \partial_\mu \big(D^\mu[A] c\big)_a \Big]\, . 
\end{eqnarray}
In order to perform the perturbative quantization  of the model and to subtract the the UV divergences, we introduce the 
 generating functional of Green's functions 
\begin{equation}\label{gen}
Z = \int \mathcal{D} X \exp\big(i S_0 + i S_{gf} + i S_{sc}\big) \,, 
\end{equation}
where $\mathcal{D} X$ is a collective notation indicating the path integral measure over the fields of the theory, while 
in the source term, $S_{sc}$, we include an external source for every field in the classical action and composite operator generated through left, 
right and BRST variation of the latter. 
For fermions and gauge bosons, we have to add sources only for the composite operators stemming from their BRST variation. 
On the other hand, every transformation acting on the scalar fields gives rise to an infinite number of composite operators as we show below.

We have shown that by acting with left and right transformations on $\subg$ and $\quoz$ two new operators appear, $\Theta^{(1,0)}$ and $\Theta^{(0,1)}$.
The variation of these quantities give rise to an infinite number of composite operators
\begin{eqnarray}
\Theta^{(n+1,m)}_{ab i_1 \dots i_n j_1 \dots j_m}:= \delta^L_{i_n} \dots \delta^L_{i_1} \, \delta^R_{j_m} \dots \delta^R_{j_1} \Theta^{(1,0)}_{ab} \,.
\end{eqnarray}
Moreover the BRST variation of $\Theta^{(n,m)}$ generates a new composite operator, namely  $\dB \Theta^{(n,m)}$. 
Using the fact that BRST transformations commute with left and right transformations, one can easily verify that the 
following source action includes all the external sources needed for the algebra of composite operators generated by left, right and BRST transformations.
\begin{eqnarray}
\!\!\!\!\!
S_{sc}&\!\!\! = &\!\!\! \Dintegral{x} 
\Bigg(\bar{J_{\psi}}\, \psi + \bar{J_{\chi}}\, \chi + \bar{{\psi}}\, J_\psi + \bar{{\chi}}\, J_\chi  + J_A^{\mu}\, A_{\mu} 
+ J_{\quoz}\, \quoz + J_{\subg}\, \subg + \sum_{\substack{n,m=0\\ n+m>0}}^{\infty} J_{scal}^{(n,m)} \Theta^{(n,m)} \nonumber\\
&&~~~~~~~~~
+\bar{\eta}\, c +\bar{c}\, \eta 
+\psi^*\, \dB \psi + \chi^*\, \dB \chi  + \bar{\psi}^*\, \dB \bar{\psi}
+ \bar{\chi}^*\, \dB \bar{\chi} + A^*_{\mu}\, \dB A^{ \mu} +c^*\, \dB c 
\nonumber\\
&&~~~~~~~~~
+J_{\quoz^*}\, \dB \quoz + J_{\subg^*}\, \dB \subg + \sum_{\substack{n,m=0 \\ n+m>0}}^{\infty} 
J_{scal^*}^{(n,m)} \dB \Theta^{(n,m)} 
+V^{\mu}\, \dB  \left(D_{\mu}\left[ A \right] \bar{c}\right)  \Bigg) \, .
\end{eqnarray}
In the above equation all internal indices have been omitted for the sake of brevity.
A conserved ghost number can be assigned by requiring that $A_{\mu}$, $\pi$, $\sigma$, $\psi$, $\chi$, $\bar{\psi}$, $\bar{\chi}$ and $b$ 
have ghost number zero, $c$ has ghost number one, $\bar{c}$, $A^*_{\mu}$, $J_{\quoz^*}$, $J_{\subg^*}$, $\psi^*$, $\chi^*$, $\bar{\psi}^*$, $\bar{\chi}^*$    
have ghost number $-1$ and finally $c^*$ has ghost number $-2$.

We define the generating functional of 1-PI Green's functions $\Gamma$, which is the Legendre 
transformation of the generating functional of connected Green's functions,
$W$ with $Z = \exp(i W)$, w.r.t. the source of the quantized fields $\psi, \ \chi,\  \bar{{\psi}},  \ \bar{{\chi}}, 
\ A_{\mu}, \ \pi,\ \sigma, \  \ c$ and  $\bar{c}$.
At tree-level the vertex functional reads
\begin{eqnarray}
\label{gamma01}
\Gamma^{(0)} &\!\!\! =&\!\!\! 
S_0 + \frac{\Lambda^{(D-4)}}{g_5^2}\,\, \dB \Dintegral{x} \left(\bar{c_a} \partial_{\mu}A^{\mu}_a \right) \nonumber \\
&&\!\!\!\!\!\!\!\!
+\Dintegral{x} 
\Bigg(\sum_{\substack{n,m=0 \\ n+m>0}}^{\infty} J_{scal}^{(n,m)} \Theta^{(n,m)} 
+\psi^*\, \dB \psi + \chi^*\, \dB \chi  + \bar{\psi}^*\, \dB \bar{\psi}
+ \bar{\chi}^*\, \dB \bar{\chi} + A^*_{\mu}\, \dB A^{ \mu} +c^*\, \dB c \nonumber\\
&&~~~~~~~
+J_{\quoz^*}\, \dB \quoz + J_{\subg^*}\,  \dB \subg + \sum_{\substack{n,m=0 \\ n+m>0}}^{\infty} J_{scal^*}^{(n,m)}\, 
\dB \Theta^{(n,m)} + V^{\mu} \,\dB \left( D_{\mu}\left[ A \right] \bar{c} \right) \Bigg)\,. 
\end{eqnarray}
The BRST invariance of eq.(\ref{gamma01}) under the BRST transformations defined in (\ref{eq.63}), (\ref{eq.64})  
can be translated into a Slavnov-Taylor (ST) identity which is to be valid 
also at the quantum level in order to fulfill unitarity. 
This implies the validity of the following  identity
\begin{eqnarray}
\label{eq:STid}
&&\!\!\!\!\!\!\!\!\!\!\!\!\!\!\!\!\!
\textit{S}(\Gamma):=\Dintegral{x}\Bigg(
\frac{\delta \QA }{\delta \psi} \frac{\delta \QA }{\delta \psi^*} 
+\frac{\delta \QA }{\delta \chi} \frac{\delta \QA }{\delta \chi^*} 
+\frac{\delta \QA }{\delta \bar{{\psi}}} \frac{\delta \QA }{\delta \bar{{\psi}}^*} 
+\frac{\delta \QA }{\delta \bar{{\chi}}} \frac{\delta \QA }{\delta \bar{{\chi}}^*} 
+\frac{\delta \QA }{\delta A^{\mu}} \frac{\delta \QA }{\delta A_{\mu}^*} 
+\frac{\delta \QA }{\delta c} \frac{\delta \QA }{\delta c^*} 
+\frac{\delta \QA }{\delta \bar{c}}  b
\nonumber\\
&&~~~~~~~~~~~
+\frac{\delta \QA }{\delta \subg} \frac{\delta \QA}{\delta J_{\subg^*}}  
+\frac{\delta \QA }{\delta \quoz} \frac{\delta \QA}{\delta J_{\quoz^*}}   
+ \sum_{\substack{n,m=0 \\ n+m>0}}^{\infty} J_{scal}^{(n,m)}  \frac{\delta \QA}{\delta J_{scal^*}^{(n,m)}}  \Bigg) = 0\,, 
\end{eqnarray}
where $\Gamma$ is the full quantum vertex functional.

\section{Local functional equation and hierarchy}
\label{sect.LFE}

The classical action given in eqs.(\ref{eq.2.14}), (\ref{Eq:gf}) has a well defined local symmetry for left and right transformations.
In what follows we provide a perturbative quantization of the model that preserves the classical symmetry, 
following the strategy devised in Refs.\cite{Bettinelli:2008qn,Ferrari:2005ii,Bettinelli:2007tq,Bettinelli:2007zn}. 
By using the invariance of the path integral measure under local left and right transformations, 
we derive two nonlinear LFEs 
that allow us to define a consistent quantum theory in the loop expansion.

The LFE associated to the local $SU(5)$ left invariance of the theory reads
\begin{eqnarray}
\label{LFE}
&&\!\!\!\!\!\!\!\!\!\!\!\!\!\!\!\!
-i\frac{\delta \QA}{\delta {\psi} }\, L_k \,\psi
-i\frac{\delta \QA}{\delta {\chi} }\, L_k \,\chi
-i \bar{\psi} \,L^k \frac{\delta \QA}{\delta \psi} 
- i\bar{\chi} \,L^k \frac{\delta \QA}{\delta \chi}   
+ \partial^{\mu} \frac{\delta \QA}{\delta A^\mu_{k}} - f^{kac} \frac{\delta \QA}{\delta A^\mu_{a}}\, A_c^{\mu}  \nonumber\\
&&\!\!\!\!\!\!\!\!\!\!\!\!\!\!\!\!
-\frac{\delta \QA}{\delta \quoz}\, \frac{\delta \QA}{\delta \big(J_{scal}^{(1,0)}\big)^k} 
-\frac{\delta \QA}{\delta \subg}\, \frac{\delta \QA}{\delta \big(J_{scal}^{(1,0)}\big)^k} 
+\sum_{\substack{n,m=0 \\ n+m>0}}^{\infty} J_{scal}^{(n,m)} \frac{\delta \QA}{\delta \big(J_{scal}^{(n+1,m)}\big)^k }
+ f^{kab} \frac{\delta \QA}{\delta c_a}\, c_b  +\nonumber\\
&&\!\!\!\!\!\!\!\!\!\!\!\!\!\!\!\!
+ f^{k a b} \frac{\delta \QA}{\delta \bar{c}_a }  \bar{c}_b 
+\psi^*\, i L^k \frac{\delta \QA}{\delta {\psi}^{*} } 
+\chi^*\, i L^k \frac{\delta \QA}{\delta {\chi}^{*} } 
+\bar{\psi}^*\,  i \frac{\delta \QA}{\delta {\bar{\psi}}^{*} }\, L^k 
+ \bar{\chi}^*\, i \frac{\delta \QA}{\delta \bar{\chi}^{*} } \,L^k 
\nonumber\\
&&\!\!\!\!\!\!\!\!\!\!\!\!\!\!\!\!
- f^{k a c}  \frac{\delta \QA}{\delta b_a } b_c 
+ f^{kab} \frac{\delta \QA}{\delta A^*_{b\mu} } A_{b\mu}
+ J_{\quoz^*} \left(\frac{\delta \QA}{\delta J_{scal^*}^{(1,0)}} \right)_{k} 
+ J_{\subg^*} \left(\frac{\delta \QA}{\delta J_{scal^*}^{(1,0)}} \right)_{k}
\nonumber\\
&&\!\!\!\!\!\!\!\!\!\!\!\!\!\!\!\!
+ \sum_{\substack{n,m=0 \\ n+m>0}}^{\infty} J_{scal^*}^{(n,m)} \left(\frac{\delta \QA}{\delta J_{scal^*}^{(n+1,m)}} \right)_{k} 
- f^{kac}  \frac{\delta \QA}{\delta c_a^{*} } c^*_b +\frac{1}{g^2}  \partial^{\mu} \frac{\delta \QA}{\delta V^{\mu}_k } 
- f^{kac}\, \frac{\delta \QA}{\delta V^{\mu}_a } V^\mu_c = 0 \, .
\end{eqnarray}
We point out that the presence of an infinite number of source terms modifies the way in which the hierarchy is realized.
This issue has been addressed  in Ref.~\cite{Ferrari:2009uj}, where it is proved that every Green's function with a finite number of external scalar fields 
depends on a finite number of ancestor amplitudes.
In this way one establishes a complete hierarchy between the 1-PI Green's functions, a fundamental aspect for the consistency of the theory.
Indeed, already at one loop level, there is an infinite number of divergent amplitudes with arbitrary number of scalar legs.
The hierarchy allows us to remove the whole set of UV divergences by means of a finite number of counterterms 
at each order in perturbation theory.

The Haar measure is also $SU_C(3) \times U_{Q} (1)$ invariant, this allows us to derive a LFE for the local right symmetry of the model
\begin{eqnarray}
\label{LFER}
&&
- \frac{\delta \QA }{\delta \subg_{\dot{a}}}  \left(\frac{\delta \QA}{\delta J_{scal}^{(0,1)}} \right)_{{\dot{a}}b} +
\left(J_{\quoz^*}\right)_{\ddot{r}} \left(\frac{\delta \QA}{\delta J_{scal^*}^{(0,1)}} \right)_{{\ddot{r}}b} 
+ \left(J_{\subg^*}\right)_{\dot{a}}  \left(\frac{\delta \QA}{\delta J_{scal^*}^{(0,1)}} \right)_{{\dot{a}}b} \nonumber\\
&&
+\sum_{\substack{n,m=0 \\ n+m>0}}^{\infty} J_{scal}^{(n,m)}  \left(\frac{\delta \QA}{\delta J_{scal}^{(n,m+1)}} \right)_{b}
+ \sum_{\substack{n,m=0 \\ n+m>0}}^{\infty} J_{scal^*}^{(n,m)} \left(\frac{\delta \QA}{\delta J_{scal^*}^{(n,m+1)}} \right)_{b} = 0\,. 
\end{eqnarray} 
The above equation has important consequences because it implies that no $\subg$ field appears at the quantum level, at least if the subtraction 
procedure is symmetric.
Indeed, if the quantum action satisfies the LFE right (\ref{LFER}), all Green's functions with one or more $\subg$ 
and no $J_{scal}$ or $J_{scal^*}$ fields vanish identically.
This can be seen as follows. Let us consider the derivative of eq.(\ref{LFER}) w.r.t. a field $\mathcal{X}$, 
with $\mathcal{X}$ different from $J_{scal}$ and $J_{scal^*}$. 
By setting the external sources to zero, we obtain
\begin{equation}\begin{split}\label{eq:a2}
\frac{\delta^{(2)} \QA }{\delta \mathcal{X}\delta \subg_{\dot{a}}}  \left(\frac{\delta \QA}{\delta J_{scal}^{(0,1)}} \right)_{{\dot{a}}b} = 0 \,.
\end{split}\end{equation}
Now, since $\left(\frac{\delta \QA}{\delta J_{scal}^{(0,1)}} \right)_{{\dot{a}}b}$ is invertible, this implies
\begin{equation}\begin{split}\label{eq:a3}
\frac{\delta^{(2)} \QA }{\delta \mathcal{X} \delta \subg_{\dot{a}}} = 0\,. 
\end{split}\end{equation}
The same procedure can be repeated for every Green's function with an arbitrary number of external fields 
different from $J_{scal}$ and $J_{scal^*}$.

We now consider the derivative of eq.(\ref{LFER}) w.r.t. $J^{(1,0)}_{scal}$
\begin{equation}
\frac{\delta^{(2)} \QA }{\delta J_{scal}^{(1,0)} \delta \subg_{\dot{a}}} \left(\frac{\delta \QA}{\delta J_{scal}^{(0,1)}} \right)_{{\dot{a}}b}
= \frac{\delta \QA}{\delta J_{scal}^{(1,1)}}\,.
\end{equation}
The above equation entails that at the tree level $ \frac{\delta^{(2)} \QA }{\delta J_{scal}^{(1,0)} \delta \subg_{\dot{a}}} $ does not vanish.
However, this kind of Green's functions do not contribute to physical S-matrix elements.
Hence, we can conclude that the $\subg$-fields do not appear in the perturbative physical spectrum of our model.

The subtraction strategy of UV divergences  is analogous to the one discussed in Refs.~\cite{Bettinelli:2008qn,Bettinelli:2007tq,Ferrari:2009uj}.
This guarantees that the counterterms are local, symmetric and depending on a finite numbers of free parameters which are 
those of the tree-level vertex functional (\ref{gamma01}).

\section{Weak power-counting}
\label{sect.wpc}

In this section we discuss a fundamental criterion in establishing which are the allowed ancestor amplitudes: 
the WPC. 
Starting from the Feynman rules provided by the tree-level quantum action (\ref{gamma01}), one can show 
that the superficial degree of divergence of a $1$-PI $n$-loop graph $G$ is bounded by
\begin{eqnarray}
\label{eq:wpc}
d\left(G\right)&\!\!\!  \leq &\!\!\!  n( D -2) + 2 -N_{A} -N_{c}-N_{\psi} -N_{\chi}-N_{\bar\psi} -N_{\bar\chi} 
-N_{V } -N_{\phi^*} \nonumber\\
&&\!\!\!\!\!
-2 \bigg( N_{J^{(n,m)}_{scal}} +N_{J^{(n,m)}_{scal^*} }+N_{c^*}+N_{\psi^*}
+ N_{\chi^*}+N_{\bar{\psi}^*}+N_{\bar{\chi}^*}+N_{ A^*}\bigg) \, .
\end{eqnarray}
where we used $N_{\mathcal{X}}$ to indicate the number of external $\mathcal{X}$ legs.
Thus at each loop order all the ancestor amplitudes, hence all the amplitudes thanks to the hierarchy, are made finite by 
a finite number of subtractions.
The proof of the formula (\ref{eq:wpc}) is straightforward in $D$ dimensions and without counterterms.
Following the procedure devised in Refs.~\cite{Bettinelli:2008qn,Ferrari:2005va}, one can show that 
the WPC bound remains valid after the introduction of the counterterms necessary 
in order to take the limit $D=4$, 
if the subtraction strategy is performed in the minimal way.
All the possible anomalous couplings that can be introduced on the basis of symmetry requirements, namely $SU(5)$ left 
and $SU(3) \times U(1)$ right invariance, are forbidden by the WPC criterion.

\section{Conclusion}
\label{sect.con}

In this paper we studied the formulation of a grand unified theory based on the nonlinearly realized gauge group. 
We focused on a particular example, namely $SU(5)$, but the main results remain valid also in the $SU(N)$ case.
Our approach is based on LFEs encoding the symmetry content  of the model and on the WPC criterion, 
which forbids the presence of anomalous couplings in the tree-level action.
The group structure forced us to use the exponential parametrization for the $SU(N)$-valued field $\Omega$ instead of the scalar parametrization adopted 
in the previous works on $SU(2)$ gauge theories.
As a consequence, the quantization  of the model requires the introduction of an infinite number of external sources which appear in both LFEs. 
Nevertheless, the Green's functions with a finite number of scalar external legs depend on a finite number of ancestor amplitudes, 
hence the theory can be consistently defined in the loopwise expansion.

In the nonlinearly realized $SU(5)$ model, we argued that a local right symmetry is a crucial ingredient in order to obtain a phenomenologically 
 successful model.
Indeed, we have shown that this right symmetry is  in a one-to-one correspondence with the unitarily implemented 
charges, i.e. the electric and the color one. 
Furthermore, at the tree-level the gauge bosons with the quantum numbers of the right generators are massless and the corresponding scalars, 
that would be physical particles, do not appear in the perturbative spectrum.
The LFE right is a fundamental tool in order to prove the absence of physical scalars in the perturbative spectrum.
Indeed, the physical S-matrix elements involving at least one scalar associated to a massless gauge boson vanish order by order in perturbation theory. 

Both the right invariance and the related LFE are novel features that are not present in the previously studied nonlinearly realized models.
As a matter of fact, the right transformations discussed here, which act only on the scalar sector and leave invariant all the other fields,  
have nothing to do with the local $U(1)$ hypercharge transformations used in references \cite{Bettinelli:2008ey,Bettinelli:2008qn,Bettinelli:2009wu}.
Indeed, the latter transformations act not only on the scalar fields, but also on fermion and gauge boson fields, depending on their hypercharge.

The phenomenologically fixed right $SU(3)\times U(1)$ symmetry reduces the number of free parameters that one can put in the tree-level action.
A detailed comparison between our model and the minimal linear one showed a significant difference in the mass structure.
In particular, one finds two independent mass parameters for the $W$ and the $Z$ bosons.
Hence, a custodial symmetry is not naturally present in the tree-level action, unless the additional mass parameter is set to zero.  

We suggested a strategy to compute the low-energy value of the Weinberg angle based on a consistent definition of the photon and $Z$ boson fields 
order by order in perturbation theory.
Furthermore, at variance with the minimal Georgi-Glashow model, the charged leptons- and the down  quarks- masses are not equal at the unified scale.
Thus, the nonlinear theory can account for the observed masses of the fermions.
The consistent formulation of a model with two very different energy scales is  a problem common to all non-supersymmetric 
grand unified theories (hierarchy problem).
In our approach to the formulation of nonlinearly realized gauge theories, 
one adopts a different point of view.
Although, in the nonlinear theory one has to impose by hand a relation between the two scales, 
the choice is stable at all orders in perturbation theory since all radiative corrections are logarithmic.

The absence of physical scalars is rather interesting especially if one considers  the doublet-triplet splitting problem that afflicts the Higgs-based grand unified theories.
Since in our model the only superheavy  particles are the $X$ and $Y$ gauge bosons, this issue seems not to be present.

The study of phenomenological aspects of grand unified theories requires to extrapolate between 
a high  and a low-energy scale. In the linear theory this is accomplished by means of  renormalization 
group techniques \cite{Georgi:1974yf}. 
The extension of such tools in the framework of  nonlinearly realized theories deserves further investigation. In this connection, 
the definition of a proper formulation for the running of the 
coupling constants seems problematic due to the lack of  multiplicative  renormalization.

\appendix

\section{Notations and conventions}
\label{app.1}

The matrix algebra $su(5)$ has 24 independent generators $L_i$, which are $5 \times 5 $ hermitian and traceless matrices.
We choose to normalize the generators so that $\text{Tr}\left( L_i L_j\right)=\frac{1}{2}\, \delta_{ij}$.
We use the following explicit representation for these matrices.
We take $L_{i}$ with $i=\{ 1, \dots 8\}$ 
\begin{equation}
L_i=\frac 12 \left( \begin{array}{c|c} \lambda_i &  0  \\ \hline 0  &  0  
\end{array} \right) ;
\end{equation}
with $ \lambda_i$ generators of $SU(3)$ in the fundamental representations, i.e. the Gell-Mann matrices.
Furthermore, for $i={9,10,11}$
\begin{equation}
L_i= \frac 12 \left(\begin{array}{c|c}
 0 &  0  \\
 \hline
 0  &  \sigma_i  
\end{array} \right) ;
\end{equation}
with $\sigma_i$ the Pauli matrices.
In particular, in the paper we used explicitly $L_{11}$ that, with our definitions, is 
\begin{equation}
L_{11}=\frac 12  \text{Diag}(0,0,0,1,-1) \ .
\end{equation}
There is another diagonal generator $L_{12}$
\begin{equation}\begin{split}
L_{12}=\frac{1}{2\sqrt{15}}\text{Diag}(-2,-2,-2,3,3) \ . \\
\end{split}\end{equation}
The remaining $12$ generators are defined using $t_a^b$, that is a $5 \times 5$ matrix with one in row $a$ and column $b$, and zero elsewhere.
We take
\begin{eqnarray}
&&
L_{12+j}=\frac 12 \left( t_4^j +t^4_j \right) \,,~~~~~~
L_{15+j}=\frac 12 \left( t_5^j +t^5_j \right)\,, \nonumber\\
&&
L_{18+j}=-\frac i2 \left( t_4^j -t^4_j \right)\,,~~~~ 
L_{21+j}=-\frac{i}{2} \left( t_5^j - t^5_j \right)\,, 
\end{eqnarray}
where  $j={1,2,3}$.
With our definitions, the $su(5)$ gauge field $A^{\mu}$ can be expressed on this basis as $ A^{\mu} = \frac 1{\sqrt 2}\, A^{\mu}_i\, L_i$.

If $T_i$ is one of the generators of the quantum symmetry, it acts, for instance, on a 
fermion field $\psi$ belonging to the fundamental representation  as 
\begin{equation}
\label{TonPsi}
\left[ T_i , \psi \right]=-L_i \psi \,.
\end{equation}
We assume that the color symmetry $SU_C(3)$ is generated by a set of  operators $T_i$ related to $L_i$ $i={1, \dots 8}$ through 
relations like those given in equation (\ref{TonPsi}).
We define the electric charge  $Q_e$ as the operator associated to the matrix 
$$\sqrt{\frac 23 } L_{\gamma}\,, ~~ {\rm with}~~ 
L_{\gamma}:=\frac 12 \left( \sqrt{\frac 32} L_{11} + \sqrt{\frac 52} L_{12}\right)\,.$$
With these definitions the most general right transformation matrix $V$ is generated by $L_a$ ($a=\{1, \dots 8 \}$) and $L_{\gamma}$.

The operator $Q_e$  gives
\begin{equation}
\left[ Q_e , \psi^j \right]=-q(j) \psi^j
\end{equation}
where $q(j)$ is the electric charge of the component $j$  of $\psi$.
This fact allows us to identify  to which component of the $SU(5)$ multiplets corresponds  each particle of the Standard Model.
The gauge field reads
\begin{equation}
A^{\mu}=
\left(
\begin{array}{ccc|cc}
G^1_1 - \frac{2 B}{\sqrt{30}} & G^2_1 & G^3_1 & \bar{X}_1  & \bar{Y}_1  \\
G^1_2& G^2_2  - \frac{2 B}{\sqrt{30}} & G^3_2 & \bar{X}_2  & \bar{Y}_2  \\
G^1_3& G^2_3 & G^3_3 - \frac{2 B}{\sqrt{30}}  & \bar{X}_3  & \bar{Y}_3 \\
\hline
X^1 & X^2 & X^3 & \frac{W^3}{\sqrt{2}} + \frac{3 B}{\sqrt{30}}  & W^+ \\
Y^1 & Y^2 & Y^3   & W^-  & -\frac{W^3}{\sqrt{2}} + \frac{3 B}{\sqrt{30}}
\end{array}
\right)\,,
\end{equation}
where $W^\pm=(W^1\mp i W^2)/\sqrt{2}$, $W^3$ and $B$ are the $SU(2)\times U(1)$ gauge bosons, while $G_{\alpha}^{\beta}$
are the $SU_C(3)$ gauge fields (gluons), with $G^\alpha_\alpha=0$.
The 12 new fields carry both $SU_C(3)$ and $SU(2)$ indices
\begin{eqnarray}
\label{eq.new}
&&
A_4^\alpha  =:  X^\alpha \qquad  (\bar{3}, 2)\,,~~~~~ 	
A_\alpha^4  =:  \bar{X}_\alpha \qquad\,\,\,  (3,\bar{2})\,,  \nonumber\\
&&
A_5^\alpha =:  Y^\alpha \qquad  (\bar{3}, 2)\,,~~~~~\,	
A_\alpha^5  =:  \bar{Y}_\alpha \qquad\,\,\,\,  (3,\bar{2}) \, ,
\end{eqnarray}
with $\alpha=1,2,3$.

Fermions in one family approximation are assigned to the following multiplets   
\begin{align}\label{eq.ferm.mult}
\bar{5}:\left(\psi^i \right)_L=\left( \begin{array}{c} d^{c1} \\ d^{c2} \\ d^{c3}  \\ e^- \\-{\nu}_e \end{array}\right)_L \\
{10}:\left(\chi_{ij}\right)_L=\frac{1}{\sqrt{2}}
\left(
\begin{array}{ccc|cc}
0  &  u^{c3}  & -u^{c2}  &  -u_1  & - d_1 \\
-u^{c3}  &  0  &  u^{c1}  &  -u_2 &  -d_2  \\
u^{c2}  &  -u^{c1}  &  0  &  -u_3  &  -d_3  \\
 \hline
u_1  &  u_2  &  u_3  & 0  &  -e^+  \\
d_1  &  d_2  &  d_3  &  e^+  &  0
\end{array}
\right)_L \, ,
\end{align}
the superscript $c$ refers to the charge conjugate of the related field.
We have chosen the phase convention in which the neutrino field appears in $\bar{5}$ (and ${5}$) with a minus sign. 
This conforms to our previous choice of $l^r =\left( \begin{array}{c} {\nu}_e  \\ e^- \end{array}\right)_L $ as a $\textbf{2}$ 
under $SU(2)$ and as being related to its conjugate $l_r =\left( \begin{array}{c} e^- \\-{\nu}_e \end{array}\right)_L $ through the antisymmetric 
tensor $l^r=\epsilon ^{rt}l_t$.

\section{Nilpotency of the BRST transformations on scalar fields}
\label{app.2}

In this appendix we show that the action of $\dB$ on the scalar fields is nilpotent.
Here we use a collective notation, $\phi$, to indicate both $\subg$ and $\quoz$.
In section \ref{sec.PI} we derived the BRST variation of the scalar fields
\begin{equation}\label{eq.ap.1}
\dB \phi_a =\Theta^{(1,0)}_{ab} c_b \ .
\end{equation}
We are going to prove that $\dB \dB \phi_a=0$.
Using  the BRST variation of the ghost fields 
\begin{equation}\label{eq.ap.3}
\dB c_a =-\frac 12 f_{abc} c_b c_c \, .
\end{equation}
one can show that 
\begin{eqnarray}
\label{eq.ap.2}
\dB \dB \phi_a = \left( \dB \Theta^{(1,0)}_{ab}\right) c_b  +\Theta^{(1,0)}_{ae} \dB c_e 
= \frac 12 \left( \Theta^{(2,0)}_{abc} - \Theta^{(2,0)}_{acb} - \Theta^{(1,0)}_{ae} f_{ecb} \right) c_c c_b \, .
\end{eqnarray}
In what follows we will show that the element between round brackets in the second equality of eq.(\ref{eq.ap.2})  
is zero. In this way we prove the nilpotency of the BRST transformations on the scalar fields.
Let $\mathcal{Q}_a$ ($a={1, \dots 24}$)  be the operators that generate the left transformations on the field, i.e.
\begin{equation}
e^{i \alpha^L_a\mathcal{Q}_a}\, \Omega\, e^{-i \alpha^L_a\mathcal{Q}_a} = e^{i \alpha^L_a L_a}\, \Omega \, ,
\end{equation}
where $\alpha^L$ is the gauge parameter.
For infinitesimal transformations this implies
\begin{equation}
\left[\alpha^L_a \mathcal{Q}_a, \phi_b \right] = \delta^L_{\alpha} \phi_b \ ,
\end{equation}
$\delta^L_{\alpha} \phi$, given in equation (\ref{variationscal}), reads
\begin{equation}
\delta^L_{\alpha}\phi_b =\Theta^{(1,0)}_{ba}\, \alpha^L_a \, .
\end{equation}
We know that the the operators $\mathcal{Q}_a$ form a representation of the $su(5)$ algebra, hence
\begin{equation}
\left[\mathcal{Q}_a , \mathcal{Q}_b \right] = f_{abc}\, \mathcal{Q}_c \, ,
\end{equation}
where $f_{abc}$ are the structure constants of the Lie algebra.
The Jacobi identity  
\begin{equation}
\left[ \mathcal{Q}_b ,\left[ \mathcal{Q}_c , \phi_a \right]\right] + 
\left[ \mathcal{Q}_c ,\left[ \phi_a \mathcal{Q}_b   \right]\right] +
\left[ \phi_a ,\left[ \mathcal{Q}_b , \mathcal{Q}_c \right]\right] = 0
\end{equation}
gives 
\begin{equation}\label{eq.J.identity}
\Theta^{(2,0)}_{abc} - \Theta^{(2,0)}_{acb} = \Theta^{(1,0)}_{ae} f_{ecb}
\end{equation}
Now, by substituting the relation (\ref{eq.J.identity}) into eq.(\ref{eq.ap.2}), we eventually obtain 
\begin{equation}
\dB \dB\, \phi_a = 0 \,.
\end{equation}

\section{Low-energy limit}
\label{app.3}

A grand unified gauge theory, in its low-energy regime, has to be well approximated by the 
Standard Model of particle physics.
In this section we study the low-energy limit of the mass sector of our model at the tree-level and prove that it is well 
approximated by the one of the  nonlinearly realized $SU(2)\times U(1)$ electroweak model \cite{Bettinelli:2008ey,Bettinelli:2008qn}.

In this appendix we adopt the following exponential parametrization for $\Omega$
\begin{equation}
\Omega=e^{i \phi }=\mathbb{I} + i \phi+\dots +\frac{\left(i \phi\right)^n}{n!}+\dots\,, ~ 
\phi=\phi^a L_a \, , \ a=1, \dots \ 24\,.
\end{equation}
The low-energy limit can be done in a more clear way if we use a gauge in which  the scalars $\phi$ associated 
to the $X$ and $Y$ gauge bosons decouple.
With this choice we can consider only those $\Omega$ that belong to the 
subgroup $SU(3)\times SU(2)\times U_Y(1)$ of $SU(5)$
\begin{equation}\label{eq.4.1}
\Omega=\left(
\begin{array}{c|c}
\Lambda & 0  \\
\hline
0 & \Sigma 
\end{array}
\right)
 \varphi \ ,
\end{equation}
where $\Lambda$, $\Sigma$ and $\varphi$ are $SU(3)$-, $SU(2)$- and $U(1)$-valued 
fields respectively. 
Moreover, it will be useful to introduce the following parametrization for $\varphi$ 
\begin{equation}\label{eq.4.2}
\varphi= \ \exp\!\Big(i \frac{\phi}{\sqrt{2}} L^{12}\Big) = \left(
\begin{array}{c|c}
\varphi_3 \mathbb{I}_3 & 0  \\
\hline
0 & \varphi_2 \mathbb{I}_2 
\end{array}
\right)
~~ {\rm with}~~ (\varphi_3)^3(\varphi_2)^2=1
\end{equation}
since $\text{det} \Omega =1 $.

From the transformation properties of $\Omega$ we derive  those of the fields 
introduced in eq.(\ref{eq.4.1}).
In the chosen gauge the transformation of $\Omega$  can be written  as
\begin{eqnarray}
\label{eq.4.4}
\Omega \rightarrow  U_3 U_2 U_1\, \Omega\,,~~~~~
\Omega \rightarrow  \Omega \, {V_3^{\dagger}} {V_1^{\dagger}}  \, ,
\end{eqnarray}
where $U_3$, $U_2$ and $U_1$ are matrices that belong to 
the $SU(3)$, $SU(2)$ and $U_Y(1)$ subgroups of $SU(5)$ 
defined in appendix~\ref{app.1}, 
while $V_3$ and $V_1$ are  matrices belonging to the $SU(3)$ and $U_Q(1)$ subgroup of $SU(5)$.
The left transformations act as
\begin{eqnarray}
\label{eq.4.5}
&&
\Lambda \rightarrow  U_3\, \Lambda\,,~~~~~  
\Sigma \rightarrow   U_2\, \Sigma\,, \nonumber\\
&& 
\varphi \rightarrow  U_1\, \varphi  \,,
\end{eqnarray}
while the right ones give
\begin{eqnarray}
\label{eq.4.6}
&&
\Lambda \rightarrow   \Lambda {V_3^{\dagger}}\,,~~~~~ 
\Sigma \rightarrow    \Sigma W_1^{\dagger}\,, \nonumber\\
&&
\varphi \rightarrow   \varphi  Y_1^{\dagger} \, ,
\end{eqnarray}
where $W_1$ and $Y_1$ are $SU(5)$ matrices with $W_1 Y_1 = V_1$. 
$W_1$ is generated by the Pauli matrix $\tau_3$  of the $SU(2)$ subgroup  
and $Y_1$ belongs the subgroup $U_Y(1)$.

Using the results derived in Sec.~\ref{sect.phys}, 
we can  substitute  $\Lambda$  in our Lagrangian  with $\mathbb{I}_3$.
Hence, from now on we use 
\begin{equation}\label{eq.4.7}
\Omega=\left(
\begin{array}{c|c}
\varphi_3\mathbb{I}_3 & 0  \\
\hline
0 & \varphi_2 \Sigma 
\end{array}
\right) \, 
\end{equation}
to study the mass Lagrangian.

Considering only one family for the sake of simplicity, one has the mass term 
\begin{equation}\begin{split}\label{eq.4.8}
(\widetilde{\psi}^{T})^{4}\, \mathcal{C}\,  \widetilde{\chi}_{4 5} = 
(\psi^{T})^{t} \,\mathcal{C} \, {\Omega}_t^4\, (\Omega^{\dagger})_4^r\, (\Omega^{\dagger})_5^u  \chi_{ru} \ .
\end{split}\end{equation}
Using equation (\ref{eq.ferm.mult})
\begin{eqnarray}
\label{eq.4.9}
 (\psi^{T})^{t}\, \mathcal{C} &= \ (\bar{e}^+_R \,,  \bar{\nu}^c_R )^t \,,~~~~~
 \chi_{ru} &= \ \epsilon_{ru}\, e^+_L
\end{eqnarray}
the term in eq.(\ref{eq.4.8}) becomes
\begin{equation}\label{eq.4.10}
(\bar{e}^+_R \,  \bar{\nu}^c_R )^t\, {\Omega}_t^4\, (\Omega^{\dagger})_4^r
\, (\Omega^{\dagger})_5^u \,\epsilon_{ru} e^+_L \ .
\end{equation}
Using the relation
\begin{equation}\label{eq.4.11}
(\Omega^{\dagger})_4^r\, (\Omega^{\dagger})_5^u\, \epsilon_{ru} = \text{det} \left( \Sigma^{\dagger} 
\varphi_2^* \right) 
\epsilon_{45}=-\left(\varphi_2^*\right)^2
\end{equation}
and substituting   $ {\Omega}_t^4$ with $\left(\Sigma \varphi_2\right)_t^4 $, equation (\ref{eq.4.10}) 
 eventually yields
\begin{equation}\label{eq.4.12}
-\left(\bar{e}^+_R \, \bar{\nu}^c_R \right)^t \left(\Sigma \, \varphi_2^* \right)_t^4 \,e^+_L \ .
\end{equation}
Proceeding as above one can show that all the mass terms   
in the action (\ref{eq.2.14})
can be written using a new field $\Phi$ instead of $\Omega$, with
\begin{equation}\label{eq.4.13}
\Phi=\left(
\begin{array}{cc}
\Sigma_4^4 \varphi_2^*  & \Sigma_4^5 \varphi_2 \\
\Sigma_5^4 \varphi_2^*  & \Sigma_5^5 \varphi_2 
\end{array}
\right) \, .
\end{equation}
Indeed, one can show that
\begin{eqnarray}
\label{eq.4.14}
&&
(\widetilde{\psi}^{T})^{\alpha}\, \mathcal{C}\,  \widetilde{\chi}_{\alpha 5} = \bar{d}_R^{\alpha}  
\big(\Sigma^{\dagger}\varphi_2^*\big)_5^t \,{\textit{q}_L}_{\alpha t}\,,~~~~~~~  
(\widetilde{\psi}^{T})^{4}\, \mathcal{C} \, \widetilde{\chi}_{4 5} = \bar{\textit{l}_R}^t 
\left(\Sigma \, \varphi_2^* \right)_t^4 {e^+_L}\,, \nonumber\\
&&
\epsilon^{5ijkl} (\widetilde{\chi}^T)_{ij}\, \mathcal{C} \,\widetilde{\chi}_{kl} = \bar{u}_R^{\alpha} 
\big(\Sigma^{\dagger} \varphi_2\big)_4^t {\textit{q}_L}_{\alpha t}\,,
\end{eqnarray}
with
\begin{equation}\label{eq.4.15}
(\textit{q}_L)_{\alpha t}  =-\left(
\begin{array}{cc}
{u_L}_{\alpha}, &{d_L}_{\alpha}
\end{array}
\right)_t \,,~~~
(\textit{l}_R)_{ t}=-\left(
\begin{array}{c}
e^+_R \\
{\nu}^c_R
\end{array}
\right)_t 
\end{equation}
Furthermore 
\begin{equation}\label{eq.4.16}
a^{t}_{5}  a^{5}_{t} +\textit{h.c.}=\text{Tr} \left(D^{\mu} {\Phi^{\dagger}} D_{\mu} \Phi \right)\,,~~~
a^{5}_{5}  a^{5}_{5} +\textit{h.c.} =\left[\text{Tr}\big(  \tau_3 {\Phi^{\dagger}} D^{\mu} {\Phi}\big) \right]^2 \,.
\end{equation}

The field $\Phi$, belonging to $SU(2)$, transforms as
\begin{eqnarray}
\label{eq.4.17}
&&
\Phi \rightarrow U \Phi \ , \, \, \, \, U \in SU(2)\,,  \nonumber\\
&&
\Phi \rightarrow  \Phi\,  \exp\Big(-i \frac{\alpha}{2} \tau^3\Big) 
\end{eqnarray}
respectively under $SU(2)$  and $U_Y(1)$ left gauge transformations, and it's invariant under right transformations.

Indeed, from our definition we have
\begin{equation}\label{eq.4.18}
\Phi^{\dagger} \Phi=\left(
\begin{array}{cc}
{\Sigma^{\dagger}}_4^4 \varphi_2 \Sigma_4^4 \varphi_2^*  + {\Sigma^{\dagger}}_4^5 \varphi_2   \Sigma_5^4 \varphi_2^*
&{\Sigma^{\dagger}}_4^4 \varphi_2 \Sigma_4^5 \varphi_2 + {\Sigma^{\dagger}}_4^5 \varphi_2  \Sigma_5^5 \varphi_2\\
{\Sigma^{\dagger}}_4^5 \varphi_2^*\Sigma_4^4 \varphi_2^* + {\Sigma^{\dagger}}_5^5 \varphi_2^*  \Sigma_5^4 \varphi_2^*
&{\Sigma^{\dagger}}_4^5 \varphi_2^* \Sigma_4^5 \varphi_2 + {\Sigma^{\dagger}}_5^5 \varphi_2^*\Sigma_5^5 \varphi_2
\end{array}
\right) = \mathbb{I}_2
\end{equation}
since $\Sigma^{\dagger} \Sigma=\mathbb{I}_2$ and $\varphi_2^*\varphi_2=1$. 
Furthermore, using the same argument, we have 
\begin{equation}\label{eq.4.19}
\text{det}\Phi = \Sigma_4^4\, \varphi_2^*\, \Sigma_5^5 \,\varphi_2 - \Sigma_5^4\, \varphi_2^*\, \Sigma_4^5\, 
\varphi_2 = \text{det}\Sigma = 1 \, .
\end{equation}
From the properties of $\Omega$ we see that, for left transformations we have
\begin{eqnarray}
\label{eq.4.20}
&&
\Sigma \rightarrow U\Sigma \ \text{ for $SU(2)$} \,,\nonumber\\
&&
\varphi_2 \rightarrow \exp\!\Big(i \frac{\alpha}{2}\Big) \varphi_2~~ \text{ for $U(1)$}
\end{eqnarray}
with $U\in SU(2)$, while for right transformations 
\begin{eqnarray}
\label{eq.4.21}
&&
\Sigma \rightarrow \Sigma \exp\!\Big(-i \frac{\beta}{2} \tau^3\Big) \,, \nonumber\\
&&
\varphi_2 \rightarrow \exp\!\Big(-i\frac{\beta}{2} \Big)\,  \varphi_2 \, .
\end{eqnarray}
These imply that for left transformations
\begin{equation}\label{eq.4.22}
\Phi \rightarrow U \Phi \exp\!\Big(-i \frac{\alpha}{2} \tau^3\Big) \ ,
\end{equation}
while $\Phi$ is invariant under right transformations.

Summing up, we argue that in the low-energy regime, our Lagrangian  can be written as a function of the Standard Model 
fermions and gauge bosons, with mass terms generated through a $SU(2)$-valued field whose transformation properties 
have been derived above.
Hence, we find exactly the mass Lagrangian proposed in Ref.~\cite{Bettinelli:2008ey}. 
Furthermore, we see that the original right $SU_C(3)\times U_Q(1)$ symmetry of the $SU(5)$ model is hidden in the 
low-energy limit. Indeed, the right transformations are absorbed within the composite field $\Phi$. 
We point out that $\Phi$ does not belong to the set of composite operators introduced for quantization of the high energy theory 
discussed in section \ref{sec.PI}. Hence, in order to discuss the low-energy limit of the model at next-to-leading order in perturbation theory, 
one has to introduce an external source for the composite operator $\Phi$.

\end{document}